\documentclass[article]{svjour3}
\usepackage{graphicx}
\usepackage{rotating}
\usepackage{amssymb}
\usepackage{mathptmx}
\usepackage{float}
\usepackage[euler]{textgreek}
\usepackage{wrapfig}
\usepackage[title]{appendix}
\usepackage{enumitem}
\usepackage{float}
\usepackage[numbers]{natbib}
\RequirePackage{xspace}
\makeatletter
\journalname{Journal of Low Temperature Physics}
\begin{document}
\title{Transition Edge Sensor Chip Design of a Modular CE\textnu NS Detector for the Ricochet Experiment }
\author{Ran Chen$^1$, H. Douglas Pinckney$^2$, Enectali Figueroa-Feliciano$^1$, Ziqing Hong$^3$, Benjamin Schmidt$^1$\\
$^1$Northwestern University\\
$^2$University of Massachusetts Amherst\\
$^3$University of Toronto}
\institute{Department of Physics, Northwestern University,\\ Evanston, IL, US \email{ranchen2025@u.northwestern.edu}}
\maketitle

\begin{abstract}

Coherent elastic neutrino-nucleus scattering (CE\textnu NS) offers a valuable approach in searching for physics beyond the Standard Model. The Ricochet experiment aims to perform a precision measurement of the CE\textnu NS spectrum at the Institut Laue–Langevin (ILL) nuclear reactor with cryogenic solid-state detectors. The experiment will employ an array of cryogenic thermal detectors, each with a mass of around 30~g and an energy threshold of 50~eV. One section of this array will contain 9 Transition Edge Sensor (TES) based calorimeters. The design will not only fulfill requirements for Ricochet, but also act as a demonstrator for future neutrino experiments that will require thousands of macroscopic detectors. In this article we present an updated TES chip design as well as performance predictions based on a numerical modeling.
\end{abstract}

\section{Introduction}
\subsection{Coherent Elastic Neutrino-Nucleus Scattering}
In 1974, Daniel Freedman predicted that for sufficiently small momentum transfers, the neutrino can interact coherently with the nucleus~\cite{joynttaillefer02}.
The process, known as Coherent Elastic Neutrino(\textnu)-Nucleus Scattering (CE\textnu NS), evaded detection until 2017, but has now been experimentally measured by the COHERENT experiment at the Spallation Neutron Source (SNS) at a neutrino energy of $\sim$~30 MeV~\cite{akimov17}.
This measurement took advantage of the fact that the CE\textnu NS cross section is roughly proportional to the square of the neutron number, which enables CE\textnu NS detectors to operate with kg-scale target mass, rather than the ton-scale target mass of conventional neutrino detectors.

\begin{figure}[!htb]
    \begin{minipage}{1.0\textwidth}
        \centering
        \includegraphics[width=1.0\linewidth, height=0.35\textheight]{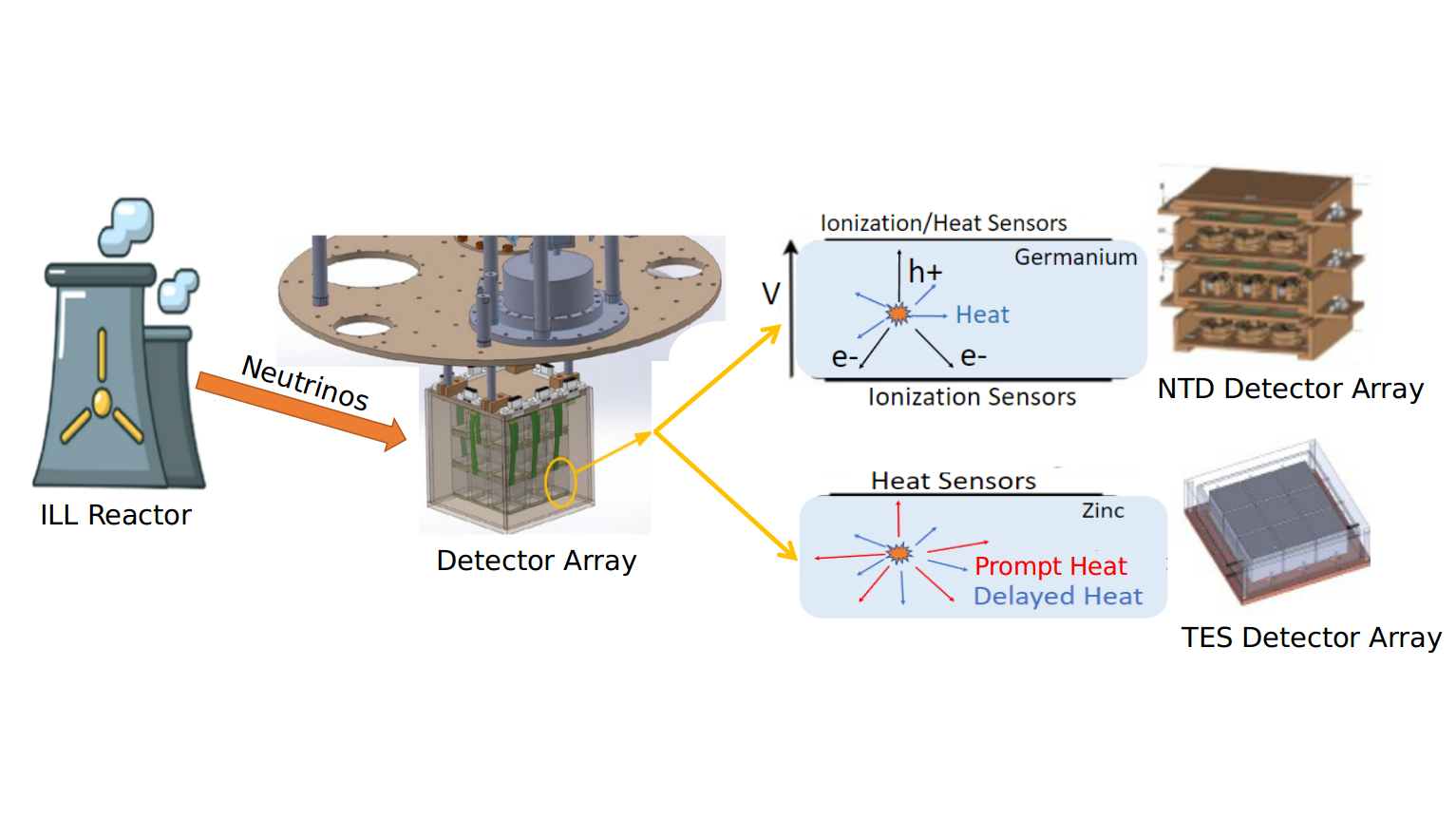}
        \caption{Ricochet Experiment Overview. The ILL Nuclear reactor provides a high intensity, low energy neutrino flux. The Ricochet experiment will deploy an array of cryogenic detectors operated at $\sim$ 10 mK to make a detailed measurement of the CE\textnu NS spectrum. The collaboration is engaged in several efforts proceeding in parallel: the CryoCube will consist of an array of 27 high purity germanium crystal NTD-based detectors with ionization and phonon readout; Q-Array will consist of 9 superconducting zinc TES-based detectors of 40 gram each.}
        \label{fig2}
    \end{minipage}
\end{figure}

\subsection{The Ricochet experiment}


The Ricochet experiment will make detailed measurements of the spectrum of CE\textnu NS at the Institut Laue–Langevin (ILL) nuclear reactor in Grenoble, France. The Ricochet collaboration will place an array of cryogenic detectors $\sim10$~m from the reactor core to perform a high-statistics, precision measurement of the CE\textnu NS spectrum.
This high-statistics measurement is made possible by the high neutrino flux generated by the ILL reactor, which is $\sim$1000 times higher than that from the SNS. However, the lower neutrino energy of reactor neutrinos requires a lower detector threshold ($\sim50$~eV) to take advantage of this increased ﬂux. To achieve such a low threshold, two technologies ---Neutron Transmutation Doped (NTD) thermistors and Transition Edge Sensors (TES)---will be used for the Ricochet experiment (Fig.~\ref{fig2}). This paper focuses on the development of the TES chip design of the superconducting detectors. 


\section{Detector Architecture}

TES-based cryogenic calorimeters achieve low energy thresholds by coupling a large mass, $\mathcal{O}(10-1000$~g), target to a thin, $\mathcal{O}(100$~nm), superconducting ﬁlm sensitive to small energy deposits.


\begin{figure}[ht]
\includegraphics[width=0.7\textwidth]{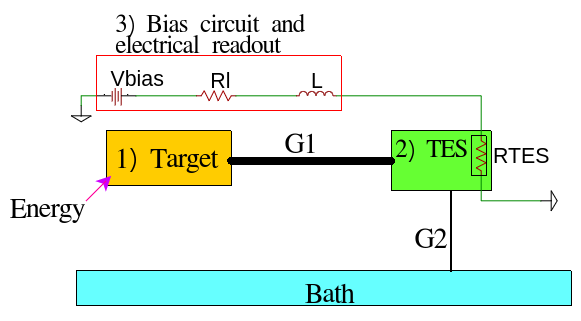}
\label{fig3}
\caption{ Diagram of a cryogenic calorimeter, including:\\
1) a particle-sensing target (eg. Si, Ge, Zn or Al);\\
2) a TES deposited on a Si chip, separate from the target;\\
3) a bias circuit and electrical readout. The TES is biased at its working point by a bias voltage $V_{\mathrm{bias}}$ across a load resistance $R_\mathrm{l}$, and coupled to a readout circuit by a inductor $L$.\\
Energy deposits in the target (1) are detected as a temperature rise and associated resistance change of the TES. The heat is conducted by G1 from target to TES (2) and then the TES translates the temperature rise into an electrical signal read by (3). Finally G2 ,which is the optimized thermal impedance, conducts the heat to the bath.}

\end{figure}

Our proposed detector architecture shown in Fig.~\ref{fig3} consists of three components: the target, the superconducting TES film deposited on a separate Si substrate and its readout. Particle interactions create a temperature rise in the target, which is held at around critical temperature of the TES in quiescent state. A thin Au film of hundreds of nm thickness is deposited on the target via evaporation or sputtering to collect that energy through electron-phonon coupling. That energy is guided from the target to the TES sensor via a Au wire bond. Finally, the heat flow out of the TES is restricted with an optimized thermal impedance called the ``meander".


This architecture allows for various target materials to be utilized in the detector, including superconducting materials like Zn and Al, and semiconducting materials like Si and Ge. Superconductors have been used in gamma-ray TES spectrometers \cite{Noroozian2013} and could provide electron/nuclear recoil discrimination with a single heat channel. This is due to different ratios of energy going into the quasiparticle and phonon systems for each type of recoil. Si and Ge targets are compatible with ionization readout for particle identification, as is being developed for the Cryocube. In addition this TES chip design is being developed with a multiplexed readout solution which would readily lend itself to a future phase 2 experiment with $\mathcal{O}$(1000) detectors for a high precision CE\textnu NS experiment as discussed in \cite{Billard_2018}.

\section{The TES Chip}
\subsection{TES Chip Design}

\begin{figure}{}
\includegraphics[width=7cm]{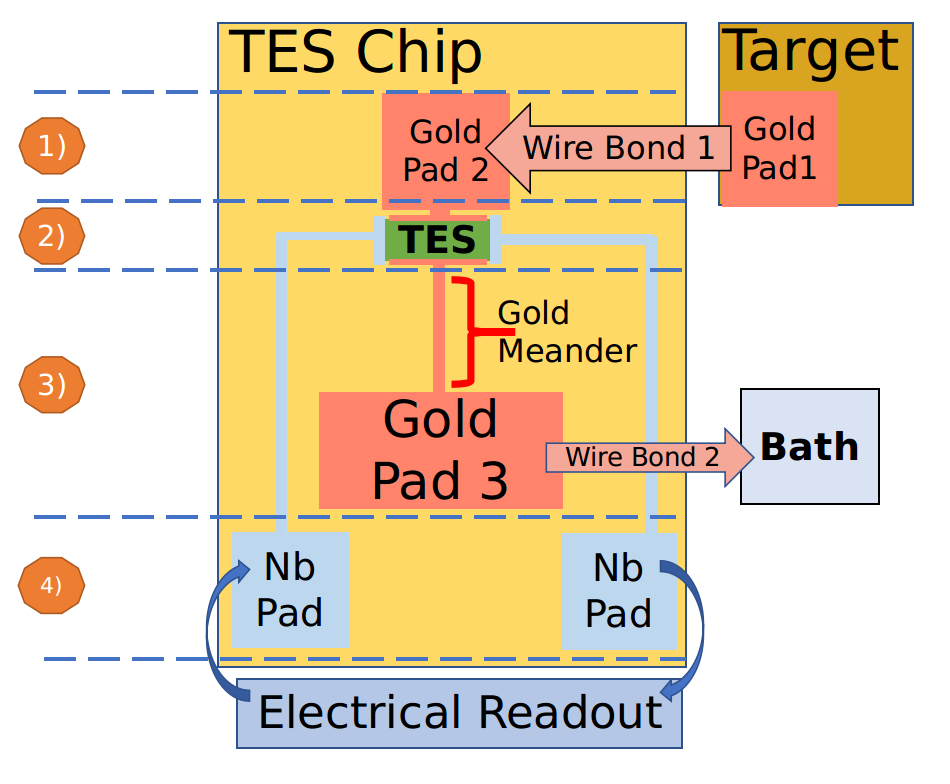}
\caption{Diagram of a TES chip design, including:\\
1) a wire bond and gold pads, to conduct the heat from the target to the TES;\\
2) a TES film, to translate temperature change into an electrical signal;\\
3) a thin gold track which is called ``meander", a gold pad, and a wire bond to the bath;\\
4) superconducting Nb traces and wire bond pads to connect the TES readout.}
\label{fig4}
\end{figure}

The TES chip outlined in Fig.~\ref{fig4} consists of 4 main components. A key aspect of which is the designed thermal impedance or “meander”.  This meander is optimized with the numerical simulation described below to obtain the best energy resolution with the pulse decay times limited to $\mathcal{O}$(10 ms) to allow for the expected event rate at ILL. A similar detector architecture was recently demonstrated in \cite{remoTES}.

\subsection{Thermal Model}

\begin{figure}[H]
    \includegraphics[width=\textwidth]{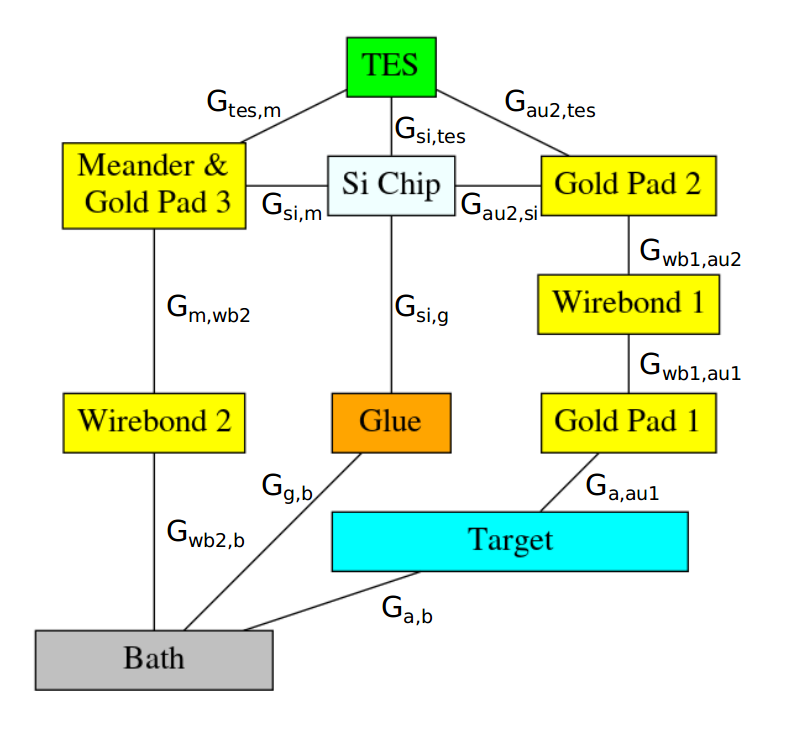}
    \caption{Block diagram representation of the thermal model of the detector design. Blocks designate different components in the detector and lines between blocks indicate their thermal connection. Abbreviations are described in Tab.~\ref{tb1}.}
    \label{fig5}
\end{figure}

\begin{table}[H]
\centering
\caption{Abbreviations used in Fig~\ref{fig5} and Tab~\ref{tab:2}}
\begin{tabular}{|l|l|l|l|l|l|} 
\hline
Abbreviation  & a        & au1       & au2         & g          & si            \\ 
\hline
Full Name  & Target & Gold Pad1 & Gold Pad2   & Glue       & Silicon Chip  \\ \hline 
\hline
Abbreviation  & tes      & m         & wb1         & wb2        & b             \\ 
\hline
Full Name  & TES      & Meander   & Wire Bond 1 & Wire Bond2 & Bath          \\
\hline
\end{tabular}
\label{tb1}
\end{table}
To optimize our design, we created a numerical model~\cite{NBastidon} which describes the detector behavior. The model simpliﬁes the detector by treating each component as a single heat capacity $C$ at a temperature $T$, connecting these components with thermal conductances $G$, as shown in Fig.~\ref{fig5}.

In the simulation, the TES response is described with a hyperbolic tangent function
\begin{equation}
R_{\mathrm{TES}} = \frac{R_\mathrm{n}}{2}\left(1+\tanh\left(\alpha\cdot\left(  \frac{T_{\mathrm{TES}}-T_\mathrm{c}}{T_\mathrm{c}}\right)+\beta \cdot \left(\frac{I_{\mathrm{TES}}-I_\mathrm{0}}{I_\mathrm{0}}\right)\right)\right),
\end{equation}
where $R_{\mathrm{TES}}$ is the TES resistance at a given operating point, $R_\mathrm{n}$ is the normal resistance of the TES, $\alpha$ is the temperature sensitivity, $T_{\mathrm{TES}}$ is the temperature of the TES, $T_\mathrm{c}$ is the critical temperature of the TES, $\beta$ is the current sensitivity, $I_{\mathrm{TES}}$ is the current through the TES, and $I_\mathrm{0}$ is the equilibrium current of the TES. The normal resistance of the TES is typically a few hundred mOhms. The TES critical temperature is assumed to be 40 mK in this study, the target for our depositions. Typical values of $I_\mathrm{0}$ are tens of uA. The sensitivity $\alpha$ and $\beta$ are assumed as 100 and 1, which are estimates based on other TES devices, for example~\cite{saab2006}.


For each part of the block diagram, the temperature evolution of element $\mathrm{i}$ is modeled as

\begin{equation}
C_\mathrm{i} \cdot \frac{dT_\mathrm{i}}{dt} = -K_{\mathrm{i,a}}\cdot(T_\mathrm{i}^{n_{\mathrm{i,a}}}-T_\mathrm{a}^{n_{\mathrm{i,a}}}) - K_{\mathrm{i,b}}\cdot(T_\mathrm{i}^{n_{\mathrm{i,b}}}-T_\mathrm{b}^{n_{\mathrm{i,b}}}) +\  \cdots,
\end{equation}
where a, b, $\ldots$ are indices of all heat capacities thermally connected to element $\mathrm{i}$, $C_\mathrm{i}$ is the heat capacity of element $\mathrm{i}$, $T_\mathrm{i}$ the temperature of element $\mathrm{i}$, and $n_{\mathrm{i,a}}$ the exponent for the appropriate thermal conduction mechanism between element $\mathrm{i}$ and $\mathrm{a}$ ($n$=2 for electron conduction, 4 for phonon conduction, and 5 for electron-phonon coupling). $K_\mathrm{i,a}$ is a constant related to the thermal conductance $G_\mathrm{i,a}$ between elements $\mathrm{i}$ and $\mathrm{a}$ by

\begin{equation}
G_{\mathrm{i,a}}(T) = \frac{dP}{dT} \biggr|_{T} = K_\mathrm{i,a} n_\mathrm{i,a} T^{n_\mathrm{i,a}-1}.
\end{equation}
Both $G$ and $K$ are scaled by the area and the length, such that

\begin{equation}
G_\mathrm{i,a} = K_\mathrm{i,a}n_\mathrm{i,a}T^{n_\mathrm{i,a}-1} = \kappa_\mathrm{i,a}\cdot A_\mathrm{i,a}/L_\mathrm{i},
\end{equation}
where $\kappa_{\mathrm{i,a}}$ is the thermal conductivity, $A_{\mathrm{i,a}}$ is the cross-sectional area and $L_{\mathrm{i}}$ is the length of element $\mathrm{i}$ along thermal conducting direction.
Heat capacities of all components are treated as temperature dependent
\begin{equation}
C = A_1\cdot T+A_3\cdot T^3,
\end{equation}
where $A_1$ and $A_3$ are material dependent heat capacity coefficients. The temperature evolution equation for the TES includes an additional term from the bias current heating

\begin{equation}
C_{\mathrm{TES}}\cdot \frac{dT_{\mathrm{TES}}}{dt} = I_{\mathrm{TES}}^2\cdot R_{\mathrm{TES}}-K_{\mathrm{TES,a}}\cdot(T_{\mathrm{TES}}^{n_{\mathrm{TES,a}}}-T_\mathrm{a}^{n_{\mathrm{TES,a}}}) - K_{\mathrm{TES,b}}\cdot(T_{\mathrm{TES}}^{n_{\mathrm{TES,b}}}-T_\mathrm{b}^{n_{\mathrm{TES,b}}}) + \ \cdots.
\end{equation}


\begin{figure}[!tbp]
  \centering
  \begin{minipage}[b]{0.8\textwidth}
    \includegraphics[width=\textwidth]{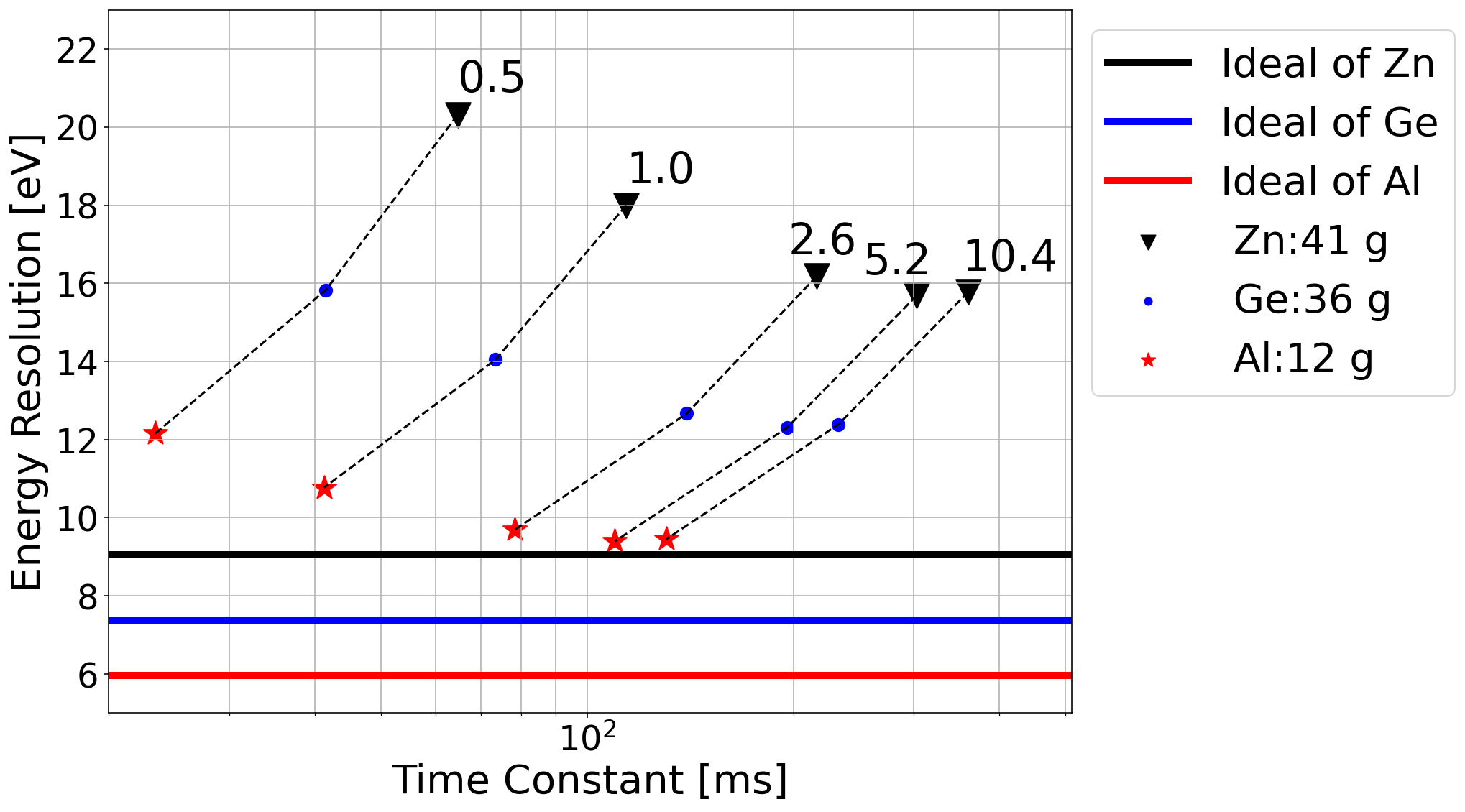}
    \caption{Simulation results of different targets and meander lengths when Tc is 40 mK. Energy resolution means baseline sigma. Numbers near points are the lengths of the meander in mm and points on the same dash line share the same meander length. Solid lines are ideal energy resolutions for devices of different target materials as outlined in Eq. ~\ref{eq8}.}
    \label{fig6}
  \end{minipage}
  \hfill
  \begin{minipage}[b]{0.7\textwidth}
    \includegraphics[width=\textwidth]{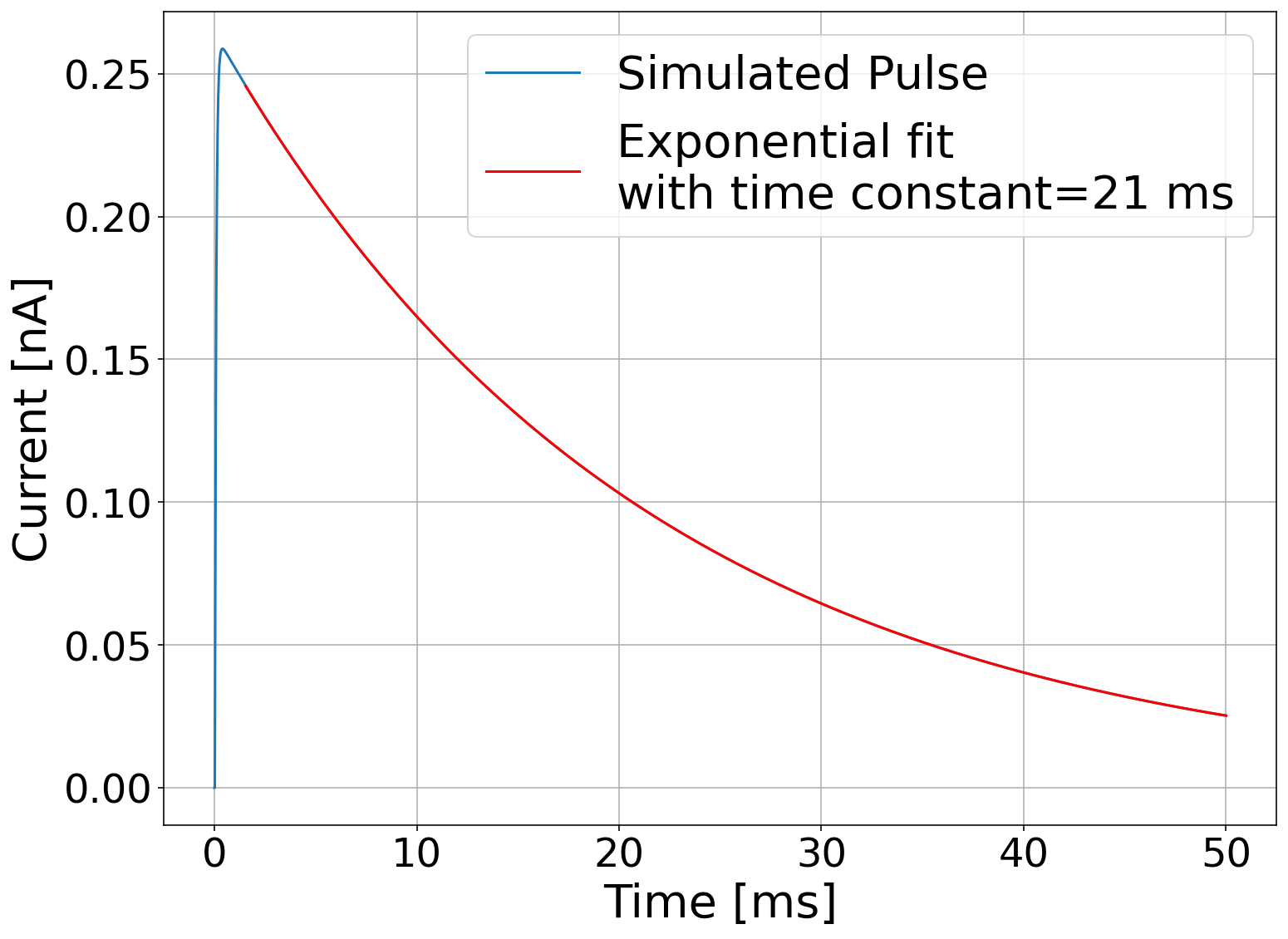}
    \caption{Simulated pulse through TES. 6 keV is input as a delta function to 12 gram Al target. The $T_c$ of the TES is 40 mK and the meander length is 0.5 mm.}
    \label{pulse}
  \end{minipage}
\end{figure}

Values for heat capacities and thermal conductances are taken both from the literature and from dedicated measurements. They are listed in Appendix A. For the key component, the Au meander  we measured the electrical resistance of a  59~mm $\times$ 50~um $\times$ 300~nm gold trace in an earlier engineering run. At 20~mK, the resistance is 26.5~$\mathrm{\Omega}$. The thermal conductivity of gold can be converted from the electrical conductivity according to the Wiedemann–Franz law:
\begin{equation}
\frac{\kappa}{\sigma} = \mathcal{L}\cdot T,
\end{equation}
where $\kappa$ is thermal conductivity, $\sigma$ is electrical conductivity, $T$ is temperature and $\mathcal{L}$ is the Lorenz number for a thin film of gold, which is taken to be  $\sim5\cdot10^{-8}~\mathrm{W\Omega K^{-2}}$~\cite{Lorentz}\cite{Wang2011}. Temperature dependent data for the Lorenz number of bulk gold is known to show a further temperature dependence with a maximum variation of up to 2 at a temperature of $\sim20$ K \cite{LorentzLowTemp}. We hence assume a remaining uncertainty of up to a factor 2, which is accounted for in the fabrication of multiple TES chips of varying meander length, see sec. \ref{sec:ChipDesign}.

In the modeling, we optimize the length of the meander, with the above numbers for different target materials, with the goal of achieving a 50 eV trigger threshold and a decay time near tens of ms. Optimization results are shown in Fig.~\ref{fig6}. Horizontal lines indicate the ideal resolution of a device with a single heat capacity equal to the total heat capacity of the system. This is calculated as \cite{sigma}
\begin{equation}
\sigma =\sqrt{\frac{4\cdot k_b\cdot T_\mathrm{c}^2\cdot C_{\mathrm{total}}}{\alpha}},
\label{eq8}
\end{equation}
where $C_{\mathrm{total}}$ is the heat capacity of the total system and $k_b$ is the Boltzmann constant.

Numbers adjacent to points in Fig.~\ref{fig6} are the meander length (in mm) for each simulation. Longer meanders result in a smaller conductance to the bath and thus slower pulse decay time constants. Slower time constants allow better thermalization of separate heat capacities and hence better approximate the ideal case scenario with a better energy resolution. If the meander is too long the conductance to the bath begins to be dominated by conduction through the TES silicon chip to the bath ($G_{\mathrm{g,Si}} + G_{\mathrm{g,b}}$ in Fig.~\ref{fig5}), bypassing the meander. After this point extra meander length only adds heat capacity and thus worsens the energy resolution. 


\section{Chip Design}
\label{sec:ChipDesign}

\begin{figure}[tb]{}
\includegraphics[width=\textwidth]{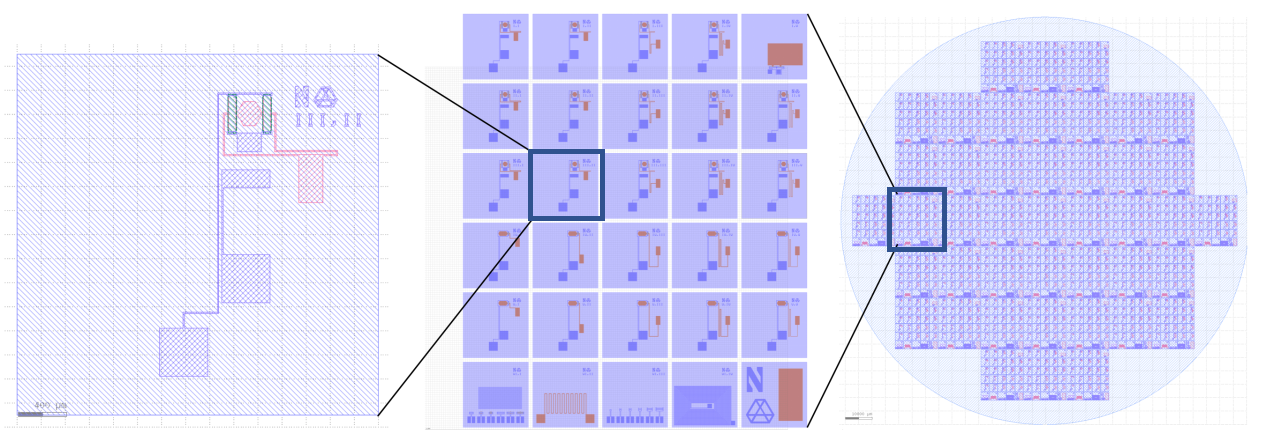}
\caption{Left: TES chip design. Gold deposition in red, Nb in blue and TES in green. Middle: A Block of TES chips. In one block, there are 25 different chips and 5 test chips. Chips are optimized for Ge, Al and Zn. Right: A 6 inch Silicon wafer with more than 1000 chips.}
\label{fig11}
\end{figure}

Fig.~\ref{fig11} shows a design with a meander length of 0.5~mm. It has two TESs in parallel to increase the heat conductance between the target and the TES which has been modeled by doubling the $G_{\mathrm{au2,tes}}$. We note that athermal TES architectures like those in \cite{Ren_2021} routinely use a large number of TESs in parallel, without reporting an anomaly in the resistive transition shape. But,  we also fabricated chips with a single TES film to perform a systematic study in the future.

In total there are 25 different TES chips packed in one block with an additional 5 test chips. We vary the meander length to attain different time constants and account for uncertainties in both fabrication and modeling of the optimal device. We further plan to test several single and double TES designs. Each chip is 3~mm$ \times$ 3~mm and more than one thousand chips can be produced on a single 6-inch wafer, allowing a consistent and efficient chip fabrication for a future large scale detector array.

\section{Conclusion}
We have created a numerical simulation of the expected performance and time response for a novel modular TES sensor architecture for cryogenic detector arrays. This simulation has been validated in a first engineering run with a previous TES chip design in 2020. In the present work it has been used to guide the design of a new iteration of TES chips and predict their expected cryogenic performance. A new wafer has been fabricated by Argonne National Laboratory and a device from this  wafer was recently tested at the University of Massachusetts Amherst with a dedicated publication in preparation. The study of quasi-particle recombination and phonon thermalization times with a superconducting Zn target and its application for particle identification is foreseen later this year. The observed long thermalization time in superconductors at low temprature \cite{Cosulich1993} could be a concern for our design. Further R\&D effort will center on measuring the effects of normal metal surface depositions on the thermalization times in different superconducting crystals.

\section{Acknowledgements}
This work is supported in part by NSF grant PHY-2013203. We would like to thank Clarence Chang and the Argonne National Laboratory team for discussions on the wafer layout, and Scott Hertel for discussions on chip layout in preparation for testing. This research was enabled in part by support provided by SciNet (www.scinethpc.ca) and the Digital Research Alliance of Canada (alliancecan.ca). This work is part of R\&D for Ricochet experiment and we would like to thank all members of Ricochet.

\section{Data Availability}
Data sharing not applicable to this article as no datasets were generated or analysed during the current study.



\begin{appendices}

\section{Parameters of the Model}

\begin{table}[ht]

\begin{center}
\begin{tabular}{ |c|c|c|c| } 
 \hline
 Component & Specific heat at 50 mK & Mass & Heat capacity \\ 
 \hline

 Ge Target & $7.17*10^{-8}$J/(K$\cdot $kg)\cite{SpecificHeat} & 36 g &  $2.60*10^{-9}$J/K\\ 
 \hline
  Al Target& $1.16*10^{-7}$ J/(K$\cdot $kg)\cite{SpecificHeat} & 12 g & $1.43*10^{-9}$ J/K  \\ 
 \hline

 Zn Target & $1.25*10^{-7}$ J/(K$\cdot $kg)\cite{SpecificHeat}& 41 g&   $5.06*10^{-9}$ J/K \\ 
 \hline
 Si Chip& $3.36*10^{-8}$ J/(K$\cdot $kg)\cite{SpecificHeat}& 8 mg& $2.82*10^{-13}$ J/K   \\
 \hline
  Au Meander(1 mm* 20 um)/Pad(1 cm$^2$) & $1.78*10^{-4}$ J/(K$\cdot $kg)\cite{SpecificHeat} & 116 ng / 0.6 mg & $2.06*10^{-14}$ J/K / $1.03*10^{-10}$ J/K \\ 
 \hline
 Glue: GE-7031 Varnish & $2.3*10^{-4}$ J/(K$\cdot $kg)\cite{pobell} & 7 mg& $1.77*10^{-9}$ J/K \\ 
 \hline
\end{tabular}
\caption{\label{tab:1}Specific heat capacities of each material in the model. Mass and heat capacity is shown for targets.}
\end{center}
\end{table}

\begin{table}[ht]
\begin{center}
\begin{tabular}{ |c|c|c| } 
 \hline
Thermal link &  Thermal conductance at 50 mK\\ 
 \hline
 G$_{\mathrm{a,b}}$ & $3.98*10^{-10}$ W/K \cite{SapphireBalls}  \\ 
\hline
 G$_{\mathrm{a,au1}}$ & $2.7*10^{-6}$ W/K\cite{Gaau1}  \\ 
\hline
 G$_{\mathrm{au1,wb1}}$  &$4.4*10^{-5}$ W/K\cite{AuConductance}  \\ 
\hline
 G$_{\mathrm{wb1,au2}}$  &$4.4*10^{-5}$ W/K\cite{AuConductance}  \\ 
\hline
 G$_{\mathrm{au2,te}}$ & $4.9*10^{-7}$ W/K*  \\ 
\hline

G$_{\mathrm{te,m}}$ & $4.9*10^{-10}$ W/K*  \\ 
\hline

G$_{\mathrm{si,te}}$ & $1.9*10^{-9}$ W/K\cite{Gaau1}  \\ 
\hline

G$_{\mathrm{si,m}}$ & $7.5*10^{-10}$ W/K\cite{Gaau1}  \\ 
\hline

G$_{\mathrm{m,wb2}}$ &$1.7*10^{-6}$ W/K\cite{AuConductance}  \\ 
\hline

G$_{\mathrm{wb2,b}}$ &$1.7*10^{-6}$ W/K\cite{AuConductance}  \\ 
\hline

 G$_{\mathrm{g,si}}$ & $1.2*10^{-6}$ W/K**  \\ 
\hline
 G$_{\mathrm{g,b}}$ & $4.1*10^{-10}$ W/K**  \\ 
\hline

\end{tabular}

\caption{\label{tab:2}Thermal conductances between each component in the model. \\* *: Calculated with the Wiedemann-Franz Law from a dedicated measurement, see equation 7.\\* **: Measured during an engineering run.}
\end{center}
\end{table}

\end{appendices}

\pagebreak
\bibliographystyle{unsrt}
\bibliography{Ricochet_TransitionEdgeSensorChipForCEvNS}

\end{document}